\documentclass{moriond}

\def\be{\begin{equation}}
\def\ee{\end{equation}}
\def\bea{\begin{eqnarray}}
\def\eea{\end{eqnarray}}

\usepackage{nameref}
\usepackage{amsmath, amsthm}
\usepackage{xspace}
\usepackage{colortbl}
\usepackage{upgreek}

\DeclareMathAlphabet{\mathpzc}{OT1}{pzc}{m}{it}
\definecolor{mygray}{gray}{0.6}
\definecolor{mauve}{rgb}{0.58,0,0.82}
\definecolor{mgreen}{rgb}{0.3, 0.85, 0.3}
\definecolor{darkgreen}{rgb}{0.1, 0.5, 0.2}
\definecolor{bred}{rgb}{0.8, 0.0, 0.0}
\definecolor{morange}{rgb}{1., 0.64, 0.}
\definecolor{darkorange}{rgb}{0.79, 0.38, 0.11}
\definecolor{pblue}{rgb}{0.2, 0.2, 0.6}
\definecolor{lightred}{rgb}{0.95, 0.6, 0.6}
\definecolor{mlightorange}{rgb}{1., 0.84, 0.}
\definecolor{lightblue}{rgb}{0.74, 0.83, 0.9}
\definecolor{lightgreen}{rgb}{0.67, 0.88, 0.69}
\definecolor{bostonuniversityred}{rgb}{0.8, 0.0, 0.0}
\definecolor{blue(pigment)}{rgb}{0.2, 0.2, 0.6}
\definecolor{ao}{rgb}{0.0, 0.5, 0.0}
\definecolor{carmine}{rgb}{0.59, 0.0, 0.09}
\definecolor{antiquebrass}{rgb}{0.8, 0.58, 0.46}
\definecolor{brown}{rgb}{0.59, 0.29, 0.0}

\newcommand{\stereo}{\textsc{Stereo}\xspace}
\newcommand{\anue}{\ensuremath{\bar{\nu}_{e}}\xspace}
\newcommand{\alphaNormU}{\ensuremath{\alpha_{l}^{\textrm{NormU}}}\xspace}
\newcommand{\alphaEscaleC}{\ensuremath{\alpha^{\textrm{EscaleC}}}\xspace}
\newcommand{\alphaEscaleU}{\ensuremath{\alpha_l^{\textrm{EscaleU}}}\xspace}
\newcommand{\ParInt}{\ensuremath{\mu}}

\newcommand{\ParIntV}{\ensuremath{\overrightarrow{\ParInt}}\xspace}
\newcommand{\Phii}{\ensuremath{\Phi_{i}}\xspace}
\newcommand{\ParNui}{\ensuremath{\alpha}\xspace}
\newcommand{\ParNuiV}{\ensuremath{\overrightarrow{\ParNui}}\xspace}
\newcommand{\MliMuAlpha}{\ensuremath{\textrm{M}_{l,i}}(\ParIntV, \ParNuiV)\xspace}
\newcommand{\Dli}{\ensuremath{\textrm{D}_{l,i}}\xspace}
\newcommand{\SEscale}{\ensuremath{S^{\textrm{Escale}}_{l,i}(\ParIntV)}\xspace}
\newcommand{\sinDeltam}{\ensuremath{\{\textrm{sin}^2(2\theta_{ee}), \Delta \textrm{m}^2_{41}\}}\xspace}
\newcommand{\ON}{\ensuremath{{\color{blue}\mathrm{ON}}}\xspace}
\newcommand{\OFF}{\ensuremath{{\color{bred}\mathrm{OFF}}}\xspace}

\newcommand{\ONacc}{\ensuremath{{\color{mygray}\textrm{ON}^{Acc}}}\xspace}
\newcommand{\OFFacc}{\ensuremath{{\color{mygray}\textrm{OFF}^{Acc}}}\xspace}

\newcommand{\GaussienneNeutrino}{\ensuremath{{\color{black}\mathcal{G_{\nu}} (A, \mu,\sigma)}}\xspace}

\newcommand{\FaccOn}{\ensuremath{\mathpzc{f}_{Acc}^{On}}\xspace}
\newcommand{\FaccOff}{\ensuremath{\mathpzc{f}_{Acc}^{Off}}\xspace}

\newcommand{\paramA}{\ensuremath{{\color{black}\textbf{a}}}\xspace}

\newcommand{\Photo}{}

\begin{document}
\vspace*{4cm}
\title{Results from the STEREO Experiment with 119 days of Reactor-on Data}

\author{ L.BERNARD, \textit{On behalf on the \stereo collaboration}\vspace{10px} }

\address{Laboratoire de Physique Subatomique de Cosmologie\\53 avenue des martyrs\\38000 Grenoble, FRANCE}

\maketitle
\abstracts{
In the past decades, short baseline neutrino oscillation studies around experimental or commercial reactor cores have revealed two anomalies. The first one is linked to the absolute flux and the second one to the spectral shape. The first anomaly, called Reactor Antineutrino Anomaly (RAA), could be explained by the introduction of a new oscillation of antineutrinos towards a sterile state of the eV mass. The \stereo detector has been taking data since the end of 2016 at 10~m from the core of the Institut Laue-Langevin research reactor, Grenoble, France. The separation of its Target volume along the neutrino propagation axis allows for measurements of the neutrino spectrum at multiple baselines, providing a clear test of an oscillation at short baseline. In this contribution, a special focus is put on the data analysis and the neutrino extraction using the Pulse Shape Discrimination observable. The results from 119 days of reactor turned on and 210 days of reactor turned off are then reported. The resulting antineutrino rate is (365.7~$\pm$~3.2) \anue /day. The test of a new oscillation towards a sterile neutrino is found to be compatible with the non-oscillation hypothesis and the best fit of the RAA is excluded at 99\% C.L.}

\section{Introduction}
\paragraph{}In the recent years, many acomplishments for neutrino physics were made close to nuclear reactors. The smallest mixing angle, $\theta_{13}$ was determined with high precision and the emitted antineutrino spectra were measured with unprecedented resolution. However, two anomalies concerning the absolute flux \cite{RAA} $-$ smaller than the prediction $-$ and the spectral shape $-$ presence of an excess around 5 MeV $-$ have yet to be solved. While they seem together to point towards a wrong prediction in the antineutrino spectra due to underestimated systematics in the measurements of the beta spectra emitted after fission \cite{Feilitzsch1982} or in the conversion method \cite{Mueller2011} \cite{Huber2011}, the first anomaly, known as the Reactor Antineutrino Anomaly (RAA), could also be solved by introducting a fourth neutrino mass eigenstate of the eV mass participating in the oscillation phenomenon.

The standard framework of 3$\nu$ mixing can be extended with the introduction of non-standard massive neutrinos. However their mixing with the active neutrinos must be sufficiently small in order not to spoil the successful 3$\nu$ mixing explanation of solar, atmospheric and long-baseline neutrino oscillation measurements. In a two flavors scheme, the survival probability of reactor electron antineutrinos depends on the antineutrino energy ($E_{\bar{\nu}_{e}}$) and its length of propagation (L) \cite{Keiser} and is written as:
\begin{equation}
P(\bar{\nu}_{e} \rightarrow \bar{\nu}_{e}) = 1 - \textrm{sin}^{2}(2\theta_{ee})\textrm{sin}^{2}(1.27\ \frac{\Delta m^{2}_{41}\textrm{[eV}^{2}]\textrm{L[m]}}{E_{\bar{\nu}_{e}}\textrm{[MeV]}})
\end{equation} 
where $\theta_{ee}$ is the mixing angle and $\Delta m^{2}_{41}$ the difference of the square of the mass eigenstates. This approximation can be used by \stereo. By 2011, global fits of short baseline neutrino oscillation data from experiments all over the world converged to a best fit for the RAA giving the following oscillation parameters: $\sin^{2}(2\theta) \simeq 0.1$ and $\Delta m^{2}\simeq 1$ eV$^{2}$  \cite{SterileNeuOscillation_globalPicture2013}. 

The RAA can be investigated with the \stereo experiment. This very short baseline reactor antineutrino experiment, running since end of 2016, is installed at 10~m from the High Flux Reactor of the Institut Laue-Langevin (ILL), in Grenoble, France. The reactor core is very compact $-$ 80 cm height and 40 cm diameter $-$ ensuring no damping of the oscillation signal. The test of an oscillation is carried out by using the segmentation of the detector into six identical cells along the neutrino propagation axis. This allows for measurements at multiple short baselines, where the neutrino propagation distance is of the order of the oscillation length. Since the oscillation probability depends on the energy of the emitted neutrino and its length of propagation, the presence of a sterile neutrino should be revealed by a distortion of the energy spectra in the different cells.

Moreover, the ILL operates with highly enriched $^{235}$U (93\%). Contributions from fission of other isotopes are thus negligible, allowing \stereo to provide a pure $^{235}$U antineutrino spectrum. This measurement will provide a test of the recent results of Daya-Bay, Reno and Double-Chooz, showing the presence of an excess of antineutrinos around 5 MeV in the measured to normalized prediction. Daya-Bay also reports an 8\% deficit in the ratio of measured to predicted flux for $^{235}$U, while no deficit is shown for $^{239}$Pu \cite{Adey:2019ywk}.

\section{Description of the Experiment}
\begin{figure}[h]
\hspace{-0.2cm}	
\begin{minipage}{.5\textwidth}
		\centering
		\hspace*{0cm}\includegraphics[width=.9\textwidth]{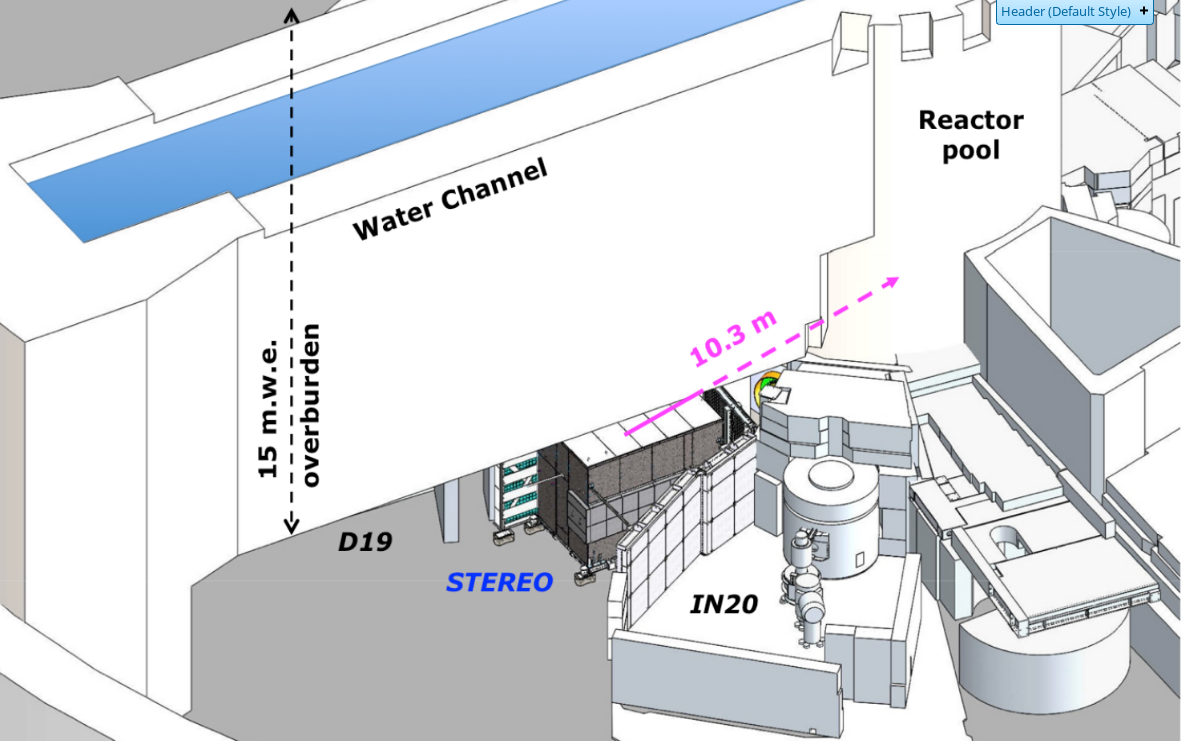}
		\caption{ILL reactor hall. The two neighboring experiments (D19 and IN20) are very close and provide a high rate of background ($\gamma$ and n). \stereo takes advantage of the 15 m.w.e overburden provided by the water channel and the reactor concrete.}
		\label{fig:ILLlocation}
	\end{minipage}%
	\hspace{10px}
	\begin{minipage}{.5\textwidth}
		\vspace{-0.45cm}
		\centering
		\includegraphics[width=.9\textwidth]{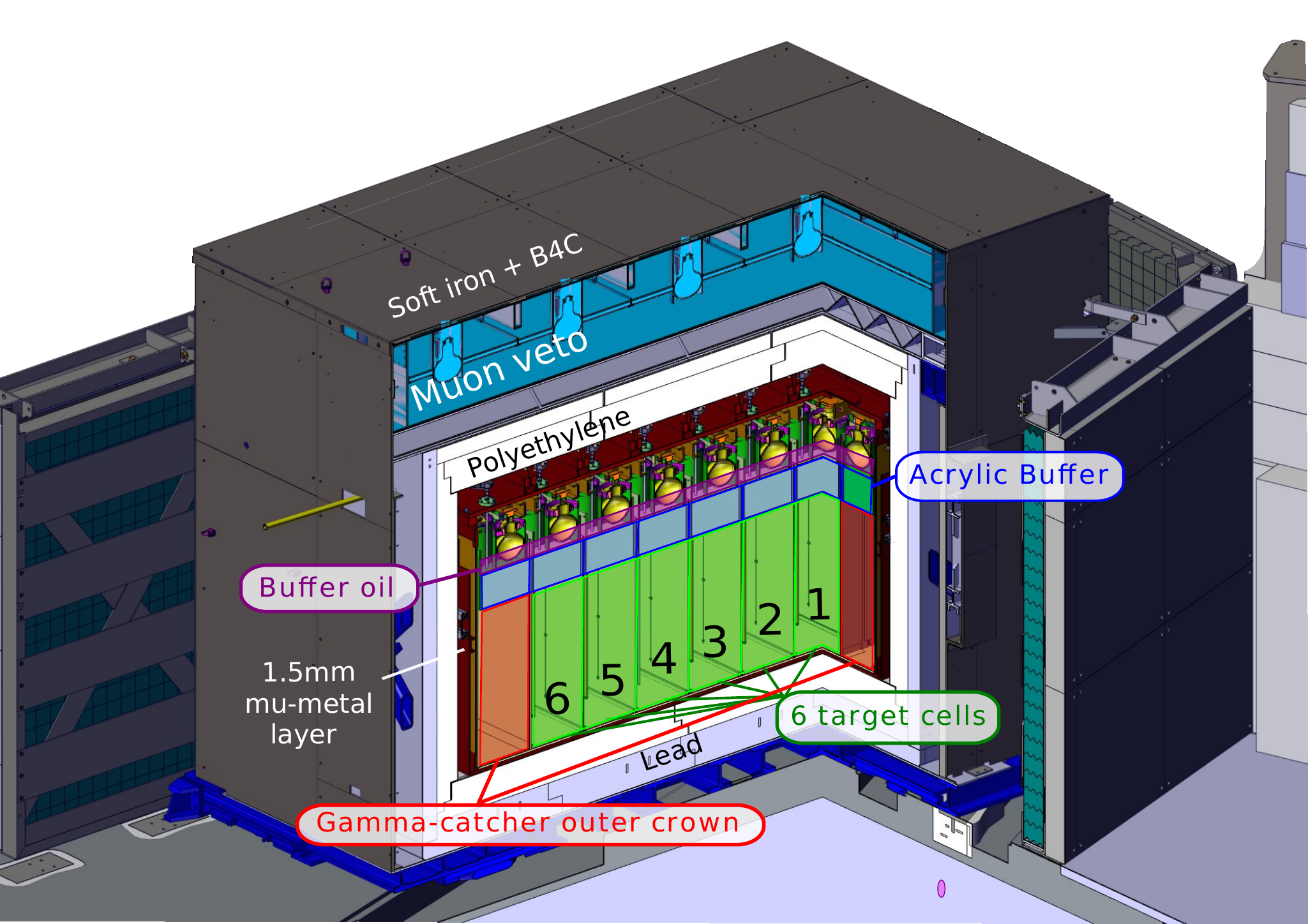}
		\caption{\stereo setup. 1-6(Green): Target cells (baselines from core: 9.4-11.1 m); (Red): two of the four gamma catcher cells surrounding the Target. The detector is enclosed in several layers of heavy shielding.}
		\label{fig:DetectorDesignAndShield}
	\end{minipage}%
\end{figure}

\paragraph{ILL reactor}The \stereo detector is located 10 m away from the reactor core, which runs at a nominal power of $\sim$58 MW$_{\textrm{th}}$. Since the reactor is mainly used as a neutron source by various experiments, the mitigation of the background generated by the neighboring experiments, shown on Figure \ref{fig:ILLlocation}, is one of the main challenges. The second challenge is related to the cosmogenic background as the \stereo detector is at sea-level. However, the overburden of concrete and water provides to the experiment a 15 m.w.e protection against this background. The essential point of the \stereo experiment is that this cosmogenic induced background can be precisely measured during the reactor stops, as the ILL operates by cycles of 50 days.
\paragraph{Detector Design and Neutrino Detection}The detector consists of an antineutrino detector, several calibration devices and a muon veto on top, all illustrated in Figure \ref{fig:DetectorDesignAndShield}. Its 2 $\textrm{m}^{3}$ Target volume segmented into 6 optically separated cells, is filled with an organic liquid scintillator (LS) where the antineutrinos are detected via inverse beta decay (IBD) on hydrogen: $\bar{\nu}_{e} + p \rightarrow n + e^{+}$. The IBD signature is a coincidence of a prompt positron and a delayed neutron capture event. The antineutrino energy is directly inferred as E$_{\anue}$ = E$_{e^+}$ - m$_{e}$ + $\Delta$M = E$_{e^+}$ + 0.782 MeV where m$_e$ is the mass of the electron and $\Delta$M the mass difference between the neutron and the proton. We will refer to the IBD events as correlated pairs. To reduce the neutron capture time, the LS has been loaded with gadolinium (Gd). The capture creates a gamma cascade with about 8 MeV total energy that can interact in the Target, but also in a surrounding crown called the Gamma-Catcher. This crown is designed to contain the $\gamma$-rays escaping from the Target and also serves as an active veto against the external background. The scintillation light is conveyed to the top of each cell by reflective walls, and read out by 48 Hammamatsu PMTs of 8 inch size. They are separated from the LS by acrylic buffers containing mineral oil that allows to establish an homogeneous light collection as well as a good optical contact between the LS and the PMTs. Only the Target LS is dopped with Gd for efficient neutron detection, confining the fiducial volume in this region. The 6 cells communicate in terms of liquid exchange, but are optically isolated to the few percent level, allowing for independent measurements of the neutrino spectrum. The detector is enclosed in a passive shielding of about 65 tons to reduce the external $\gamma$ rays and neutron background. A water Cerenkov tank, placed on the top of the inner detector, acts as an active shielding by vetoing all muons crossing its volume. The deposited light is read by 20 PMTs. More details are given in \cite{Allemandou:2018vwb}. 

\begin{figure}[!h]
	\centering
	\hspace*{-0.8cm}
	\includegraphics[width=1.065\textwidth]{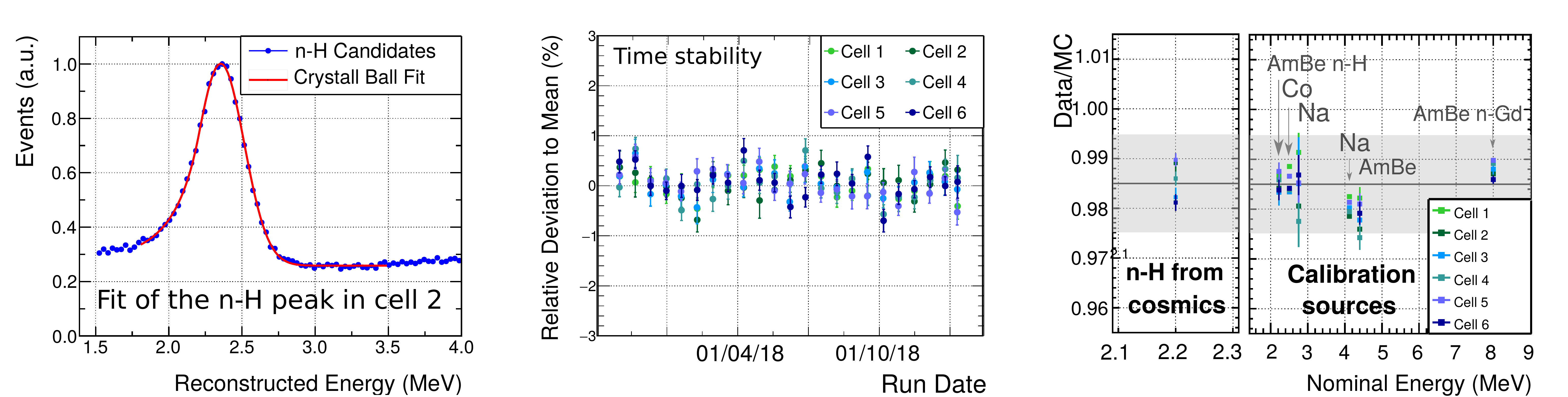}
	\caption{Evaluation of the systematic uncertainties associated to the reconstruction of the energy method. The peak from the capture of neutrons on hydrogen is fitted for each cell and each energy bin. Its time stability is monitored. We find that a 1.5\% residual discrepancy is present between data and MC, but well described by the use of several calibration sources. We have used a rescaling for the present analysis but the origin of this discrepency has been understood and there will be no need for rescaling in future results.}	
\label{fig:nH_Peak_Cell2_custom}
\end{figure}

\paragraph{Simulation} A Geant4 (version 10.2) Monte Carlo model (MC) based on DCGLG4sim describes the detector geometry, the shielding, and the position to the reactor core. It also includes particle interactions including neutron moderation and capture; light production, transport taking into account cross talks between cells, detection, and signal conversion in the electronics. The MC simulation is implemented in such a way that the output is given in the same format as the real data.
\paragraph{Detector Response}An automatic and daily monitoring of the electronic and liquid stability is done using light pulses from LED. A set of $\gamma$ and neutron sources are regularly placed inside, below and around the detector to ensure the monitoring of the energy response. A dedicated algorithm to reconstruct the deposited energy from the collected light has been developed, taking into account the evolution of the light cross-talks and the light collection in each cell along time \cite{Allemandou:2018vwb}. To evaluate the systematics associated to this method, we use the cosmic induced n-H peak since its quasi-uniform distribution is similar to the one of the neutrinos. Figure \ref{fig:nH_Peak_Cell2_custom} shows the fit of the n-H peak of cell 2 and its time stability for the 6 cells. The comparison of the reconstructed energy of different sources, as well as the n-H peak, with the simulation is then used to rescale the Monte-Carlo simulation from a 1.5\% residual discrepancy. This difference $-$ originating from the removal of very low PMT charges $-$ is however understood and will be corrected in the next analyses of \stereo. The systematic uncertainties are derived from the studies of the n-H peak. Their value are given in the oscillation analysis (cf. Section \ref{sectionOscillationTest}).

\section{Signal Selection}
\paragraph{}To select the neutrino candidates, a set of cuts corresponding to the best compromise between detection efficiency and background rejection is applied. Beyond the basic cuts on energy and capture time of the prompt and the delayed events, the detector segmentation is exploited to tag the topology of energy deposition of IBD events. The prompt and delayed events have to take place in a time period of less than 70 $\upmu$s and within a distance of 600 mm. We require for the total energy of the prompt event to be within 1.625 MeV and 7.125 MeV. This range ensures to keep the majority of the neutrino signal while excluding the low-energy background contamination. From the characteritics of the energy deposits of the positron signal, less than 1 MeV must be deposited in the neighboring cells and less than 0.4 MeV in the other cells. This allows for the two 511 keV $\gamma$ rays to escape along with some light leaks. The delayed event must have an energy between 4.5 and 10 MeV, while at least 1 MeV has to be detected in the target, allowing to reject high energy background events coming from the sides into the detector. The cosmogenic background, such as muons or fast neutrons coming from the muon spallation in the shieldings is reduced by applying a veto of 100 $\upmu$s after the identification of a muon in the Cerenkov water tank or in the Gamma-Catchers. An isolation of 100 $\upmu$s before the prompt and after the delayed is also required. Finally, the muons stopping in the upper part of the detector can be identified by their asymmetry of light collection on the 4 PMTs of the cell. 
The mean efficiency of the above selection cuts is evaluated using millions of neutrinos generated in the \stereo simulation. Its value is (61.4 $\pm$ 0.4)\% where the uncertainty is estimated from the propagation of the uncertainty on the energy scale only.

The dominant uncertainty in the efficiency comes from the neutron detection efficiency (delayed event). The evaluation of this uncertainty is carried out with dedicated studies using an americium-beryllium source deployed inside and around the detector. The emitted neutron is captured on the Gd nuclei present in the LS, allowing to compute the fraction of neutron captures in both the data and the simulation. The thermalization and neutron capture time are well reproduced by the simulation. As the Gamma-Catcher is not loaded with Gd, the Gd-fraction decreases towards the edges of the Target volume. A three-dimensional model is used to describe the local value of Gd-fraction, providing a 3D map of the data/MC model. This accurate measurement allows for a correction associated to a sub-percent level systematic uncertainty ($\pm$0.91\%).
%
%
%
\section{Backgrounds}The accidental background, i.e. the selection of two uncorrelated events, is mainly discriminated by the 16 $\upmu$s capture time constant of the neutron capture and the prompt-delayed distance. Its precise measurement for each cell and energy bin is carried out by opening successive time windows ($\sim 10$) of 70 $\upmu$s, spaced by 1 ms and starting from the prompt event. The IBD cuts are applied and eventually, a delayed event is found in the window, forming an accidental pair. The accidental background is strongly linked to the activity of the neighboring experiments. 80\% of the total rate of accidental IBD candidates ($\sim$ 130 events per day) is contained under 3 MeV. The remaining correlated background, i.e. two events coming from the same physical process, can only be measured during the phases where the reactor is turned off. It is dominated by two categories of cosmic induced events. A fast neutron produced from the spallation of muons in the heavy shielding can penetrate the detector, inducing a prompt nuclear recoil signal in the right energy window, before being captured on a Gd nucleus of the LS. Two fast neutrons coming from the same cosmic shower might also punch through the shielding, thermalize in the liquid and get captured in the Target to form the prompt and delayed signals. A third feature, arising around 5-6 MeV in the correlated background spectrum comes from the $^{12}$C(n,n'$\gamma$)$^{12}$C reaction in the LS.

\section{Neutrino Spectra Extraction using the PSD Parameter}
\label{sectionNuExtraction}	
\begin{figure}[h!]
	\centering
\hspace{-1.3cm}\includegraphics[width=.9\textwidth]{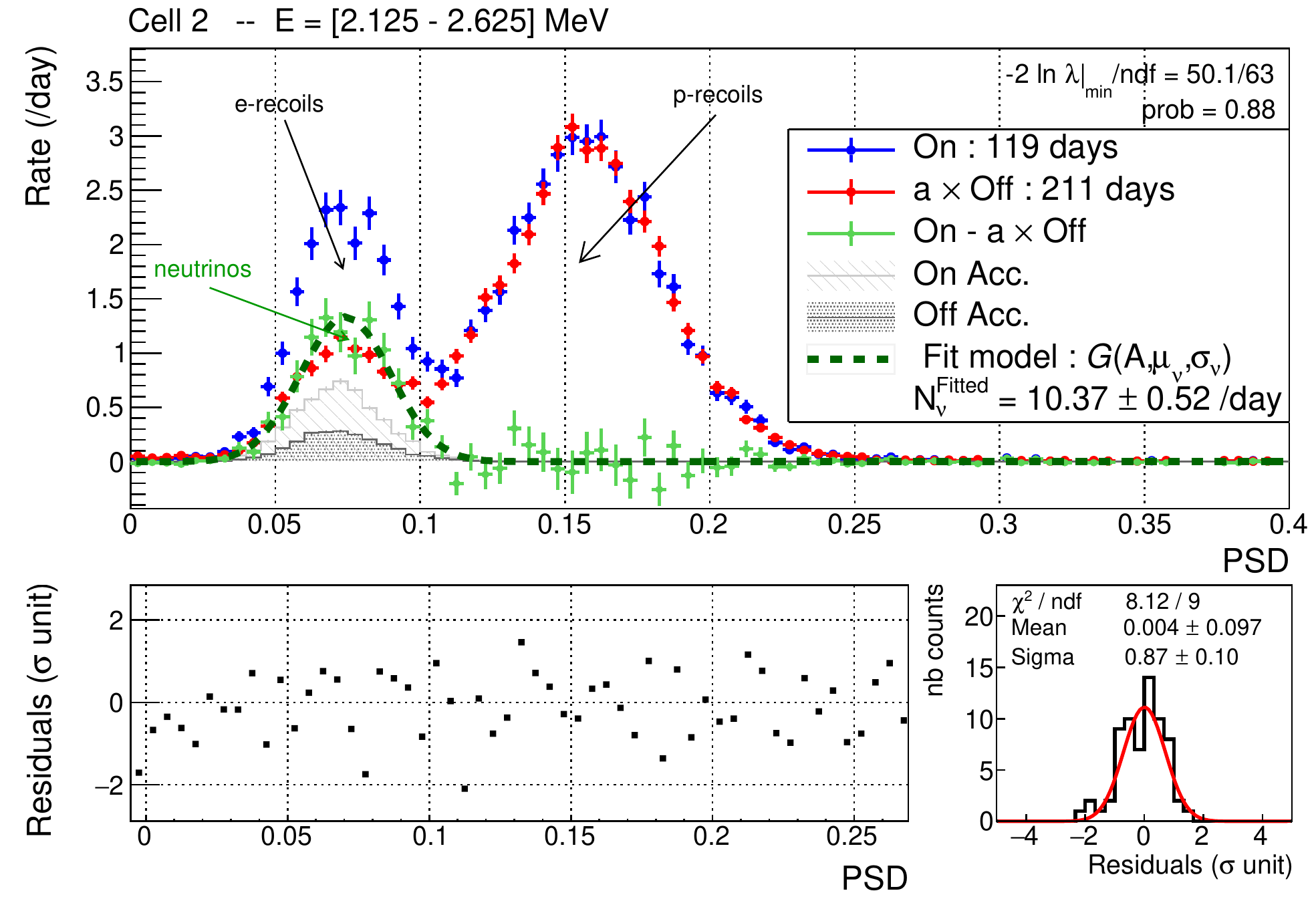}
	\caption{Example of neutrino extraction for one bin and one cell using the PSD parameter. The PSD figure is made of two main components: the electronic recoils and the proton recoils. The measured reactor-off correlated pairs (red), combined with the accidental pairs (light gray and dark gray), provides the background model of the reactor-on correlated pairs (on). The neutrinos are modeled by a Gaussian (dark green) and superimpose to the "On-\paramA$\times$Off" under the electronic component. A parameter \paramA accounts for any normalization difference between the reactor-off and the reactor-on periods, such as acquisition time or pressure variations. The minimization is done using the inverse of a ratio of likelihood $-2\ln\lambda$, latter follows a $\chi^2$ law. The quality of the fit is given by the probability (0.88). The residuals (bottom left plot) demonstrate no systematic effect in the shape of the distribution. Their distribution (bottom right plot) follows a normal law.}
\label{fig:PSDFigures_Cell2_Ebin3}
\end{figure}

\begin{figure}[h!]
	\hspace{-0.3cm}
	\begin{minipage}{0.48\textwidth}
		\centering
		\hspace*{-0.4cm}\includegraphics[width=1.1\textwidth]{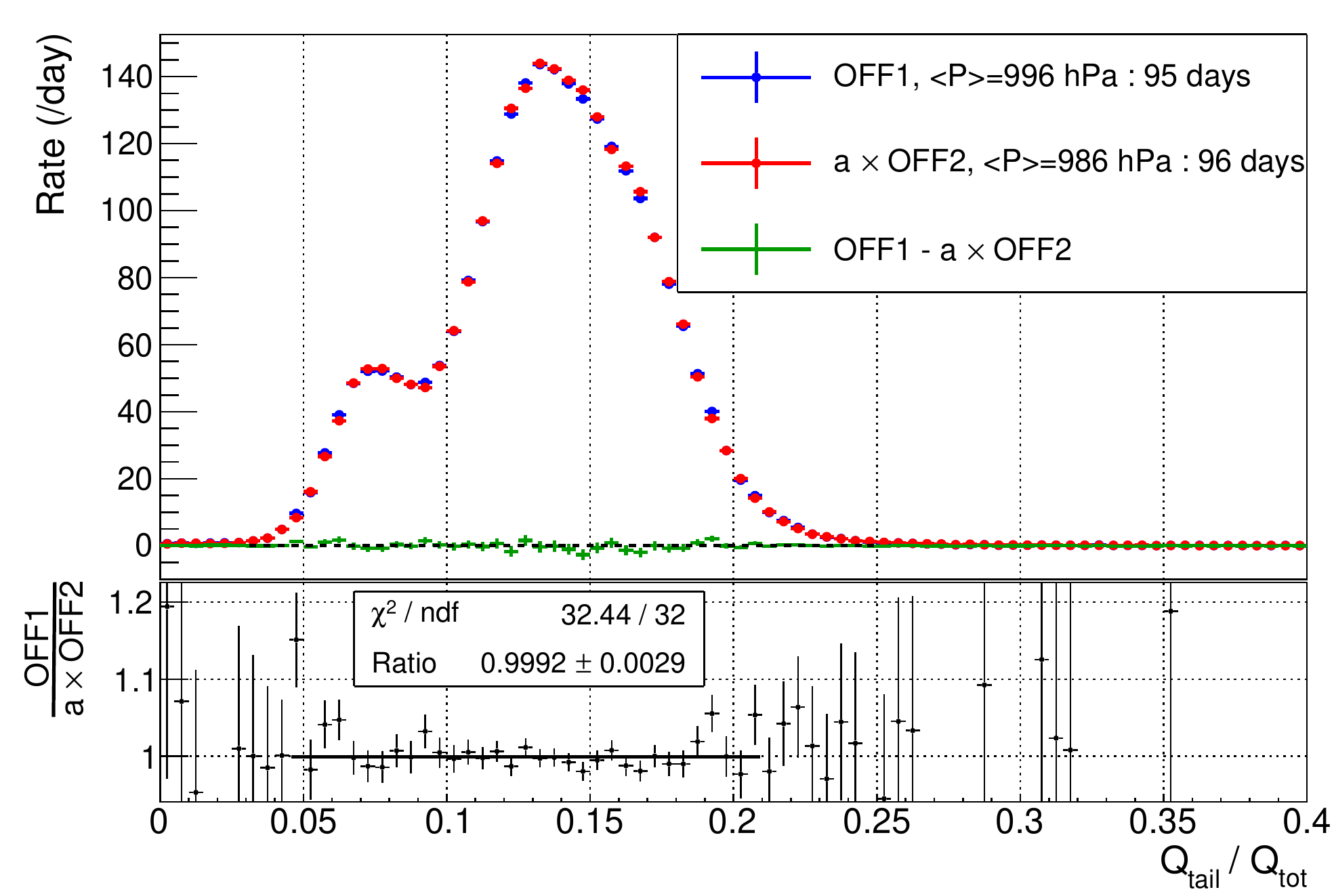}
	\end{minipage}%
\hspace*{1.cm}
	\begin{minipage}{0.48\textwidth}
	\centering
	\hspace{-0.9cm}\includegraphics[width=1.1\textwidth]{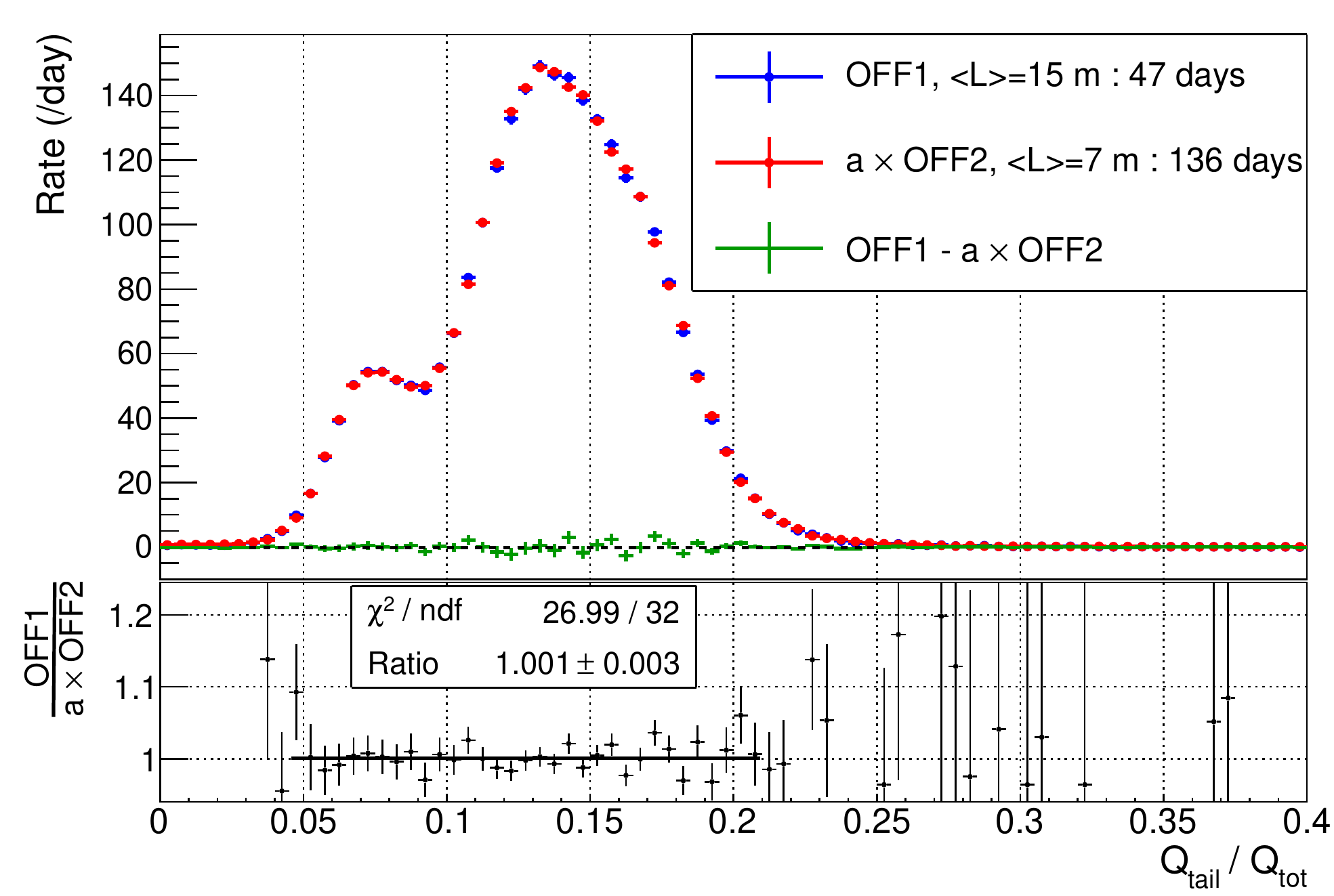}
\end{minipage}
	\caption{Test of the background shape stability under pressure variation (left) and variation of the level of water in the reactor pool (right). The whole reactor-off period is splitted in two bins containing the high and low pressures. The residuals between the two populations can be zeroed with a single free normalization parameter "a", demonstrating a high stable shape of the background PSD.}
	\label{fig:PSDfigures_pressure_Target_custom}
\end{figure}

\paragraph{Principle of the Method} 
The Pulse Shape Discrimination (PSD) properties of the LS allow to disciminate neutrinos from the dominant remaining cosmic background. This observable is defined as the ratio of the pulse tail over the total charge. Its evolution in time, mainly caused by temperature variations, is monitored and corrected using the accurate PSD mean value of the single events as a monitor of the daily drifts. The PSD distribution of all correlated prompt events after selection cuts is displayed for one cell and energy bin in the Figure \ref{fig:PSDFigures_Cell2_Ebin3}, for the periods of reactor-off and reactor-on. The distributions of accidentals pairs are also used in the neutrino extraction. The PSD distribution of the background is divided into an electronic recoils part at low PSD (captures of fast neutrons on H or Gd or accidentals) and a proton recoils component at high PSD (induced by fast neutrons in the LS). The correlated IBD candidates of the reactor-off periods (\OFF) and the accidental components (\OFFacc and \ONacc), provide a background model that is used to fit the correlated IBD candidates (\ON) using a log-likelihood maximization, required by the low statistics of some PSD bins. A Gaussian component (\GaussienneNeutrino) let totally free in the maximization is used to describe the distribution of the neutrinos. The correlated reactor-on data is modeled as follow:
\begin{equation}
\ON = \paramA \times (\OFF - \FaccOff\ \OFFacc) + \FaccOn\ \ONacc + \GaussienneNeutrino 
\end{equation} where the parameter \paramA takes into account the difference of normalization between the two periods that is caused by distinct acquisition time and other environemental variables detailed in the paragraph \textit{\nameref{sectionBackgStab}}. \FaccOff and \FaccOn are correction factors used to account for the different dead-time calculation of an accidental and a correlated pair. The neutrino rates are extracted for each cell and energy bin of size 500 keV, providing 6 spectra that are displayed in the Figure \ref{fig:Ledieb_Spectra2}.
\paragraph{Monte-Carlo Simulation} \label{sectionMCNuextraction} A Monte-Carlo simulation based on 5000 pseudo-experiments is developed to evaluate the bias of the likelihood estimator at low statistics. The background model is derived from the real off-periods PSD distributions and also takes into account the accidental components. The bias is negligible across most of the energy range and reaches a maximum value of 1.7\% at 7.125 MeV. It is used to correct the number of IBD candidates. The probability density functions of the minimized ratio of likelihood $-2 \ln\lambda$ are also built and demonstrated to follow $\chi^2$ laws with $n$ degrees of freedom, providing the statistical test that is used to quantify the quality of each fit.
\paragraph{Background Stability} 
\label{sectionBackgStab}
The only hypothesis on which this method relies is that the background shape stays constant from the reactor-off to reactor-on periods. The variations of the cosmic background are driven by two parameters having different mean values between the reactor-off and reactor-on periods: the atmospheric pressure ($\Delta P_{\textrm{ON}\rightarrow\textrm{OFF}} \sim$~10~hPa) and the level of water laying in the reactor pool sitting on top the core ($\Delta L_{\textrm{ON}\rightarrow\textrm{OFF}} \sim$~8~m) (cf. Figure \ref{fig:ILLlocation}). The PSD shape stability is tested for these two variables by splitting the reactor-off periods in two bins gathering the IBD candidates obtained under high pressures (water level) and under low pressures (water level). The reactor-off time offers the possibility to study the effect of the water level, with 136 days of empty pool $<L>$=7~m and 47~days of filled pull $<L>$=15~m. These high and low contributions are displayed on Figure~\ref{fig:PSDfigures_pressure_Target_custom} and the residuals are plotted, demonstrating a very stable shape of the PSD and validating the method with high precision.

\section{Oscillation Hypothesis Testing}
\label{sectionOscillationTest}
\paragraph{}The dataset gathers 119 days of reactor-on and 210 days of reactor-off, between November 15, 2017 and January 15, 2019. It is splitted in energy bins of 500 keV between 1.625 MeV and 7.125 MeV. By summing the 6 neutrino spectra, we obtain a total rate of 365.7 $\pm$ 3.2 \anue/day. 
\paragraph{Formalism}The oscillation analysis is done in an independant way from any reactor prediction. It tests the shape of the neutrino spectrum by comparing the measured number of IBD candidates in each bin $i$ of cell $l$ \Dli to their respective expectation $M_{l,i}$. The parameters of interest \ParIntV are \sinDeltam. To take into account the systematic uncertainties, nuisance parameters \ParNuiV are added to the model, allowing more flexibility to adapt to the data:
\begin{equation}
\MliMuAlpha = M_{l,i}(\ParIntV) \ \left(1 \ + \ \alphaNormU + \ \SEscale\ .\ (\alphaEscaleC + \alphaEscaleU) \right)
\end{equation} 
The \ParNuiV include the uncorrelated uncertainty on the normalization (cell volume (0.83\%) and neutron efficiency corrections (0.91\%)), the uncorrelated energy scale uncertainty (Mn anchor point (0.2\%) and cell-to-cell deviations (0.5\%)), and the correlated energy scale uncertainty (time stability (0.3\%) and anchoring of the Mn(1\%)). The sensitivity factors \SEscale describe the variation of the number of IBD candidates under energy scale variation. A free parameter \Phii is introduced for each energy bin but common to all cells. This way, the choice of the initial predicted spectrum becomes arbitrary and the best spectrum shape common to the 6 cells is determined independently of a reactor prediction. A $\Delta\chi^2$ method is used to test the different oscillation hypotheses of the parameter space. For each hypothesis \sinDeltam, 10000 pseudo-experiments allow to construct the $\Delta\chi^2$ probability density functions that are later used for the decision. The $\chi^2$ is expressed as:
\begin{align}
\label{eq:Chi2oscillation}
\chi^2 = &\sum_{l=1}^{N_{\text{Cells}}} \sum_{i=1}^{N_{\text{Ebins}}} \left(\, \frac{\Dli - \Phii\ \MliMuAlpha}{\sigma_{l,i}} \right)^{2} 
\\\nonumber + & \sum_{l=1}^{N_{\text{Cells}}} \left(\frac{\alphaNormU}{\sigma_l^{\text{NormU}}}\right)^2 +
\left(\frac{\alphaEscaleC}{\sigma^{\text{EscaleC}}}\right)^2 +
\sum_{l=1}^{N_{\text{Cells}}} \left(\frac{\alphaEscaleU}{\sigma_l^{\text{EscaleU}}}\right)^2
\end{align}
where the $\sigma^X$ are constraining terms associated to the nuisance parameters coming from the uncertainty studies. The $\sigma_{l,i}$ is the statistical uncertainty on the model \MliMuAlpha tested at the current point of the parameter space. It can deviate significantly from the statistical uncertainty of the data when large mixing angles are tested, as illustrated in the next section.

\begin{figure}[h!]
	\centering\includegraphics[width=.65\textwidth]{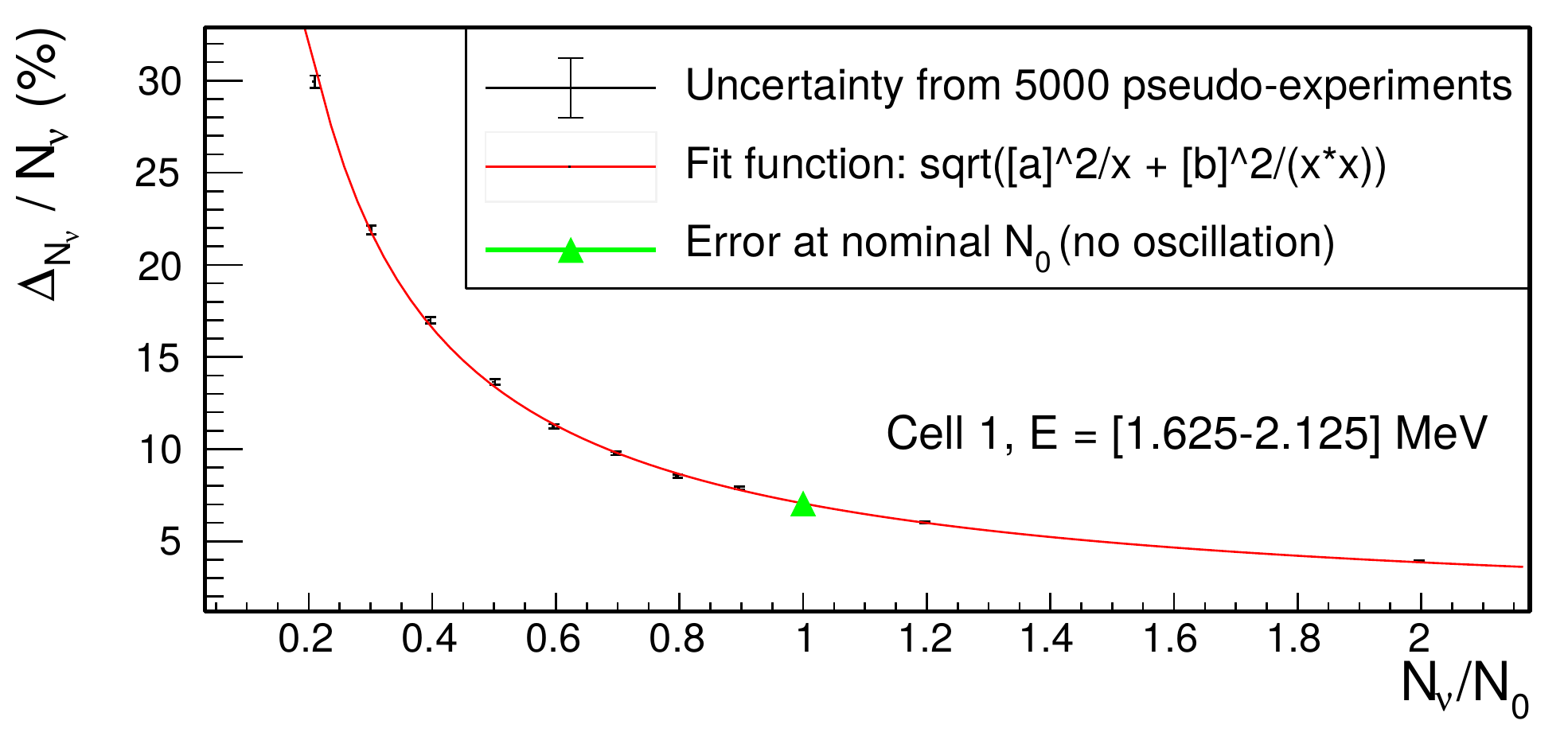}
	\caption{Statistical uncertainty of the neutrino rate prediction for cell 1 at 1.875 MeV. }
	\label{fig:Cell1Ebin0_proceeding}
\end{figure}

\paragraph{Uncertainty Prediction} To obtain the $\sigma_{l,i}$, the simulation described in the paragraph \textit{\nameref{sectionMCNuextraction}} is used. The rate of generated neutrinos $N_\nu$ is varied from 20\% to 200\% of a nominal expected number of neutrinos $N_0=370\ \anue$/day, leading to the black points on the Figure \ref{fig:Cell1Ebin0_proceeding}. We use $x$ to denote the ratio $N_\nu/N_0$. Since the statistical uncertainty depends on the signal statistics $-$ evolving as $1/\sqrt{x}$ $-$ and the constant background statistic $-$ evolving as $1/x$ $-$ , we make use of the following fit function:
\begin{equation}
\frac{\Delta{N_{\nu}}}{N_{\nu}} = \sqrt{ a^{2} \times \frac{1}{x} + b^{2} \times \frac{1}{x^{2}}}
\end{equation}
where the coefficients $a$ and $b$ allow to adapt to the signal-to-background ratio of a given couple \{cell, energy\}. Figure \ref{fig:Cell1Ebin0_proceeding} shows such an adjustment for one cell and one energy bin.

\begin{figure}[h]
	\begin{minipage}{0.58\textwidth}
		\centering
		\hspace{-.8cm}\includegraphics[width=.95\textwidth]{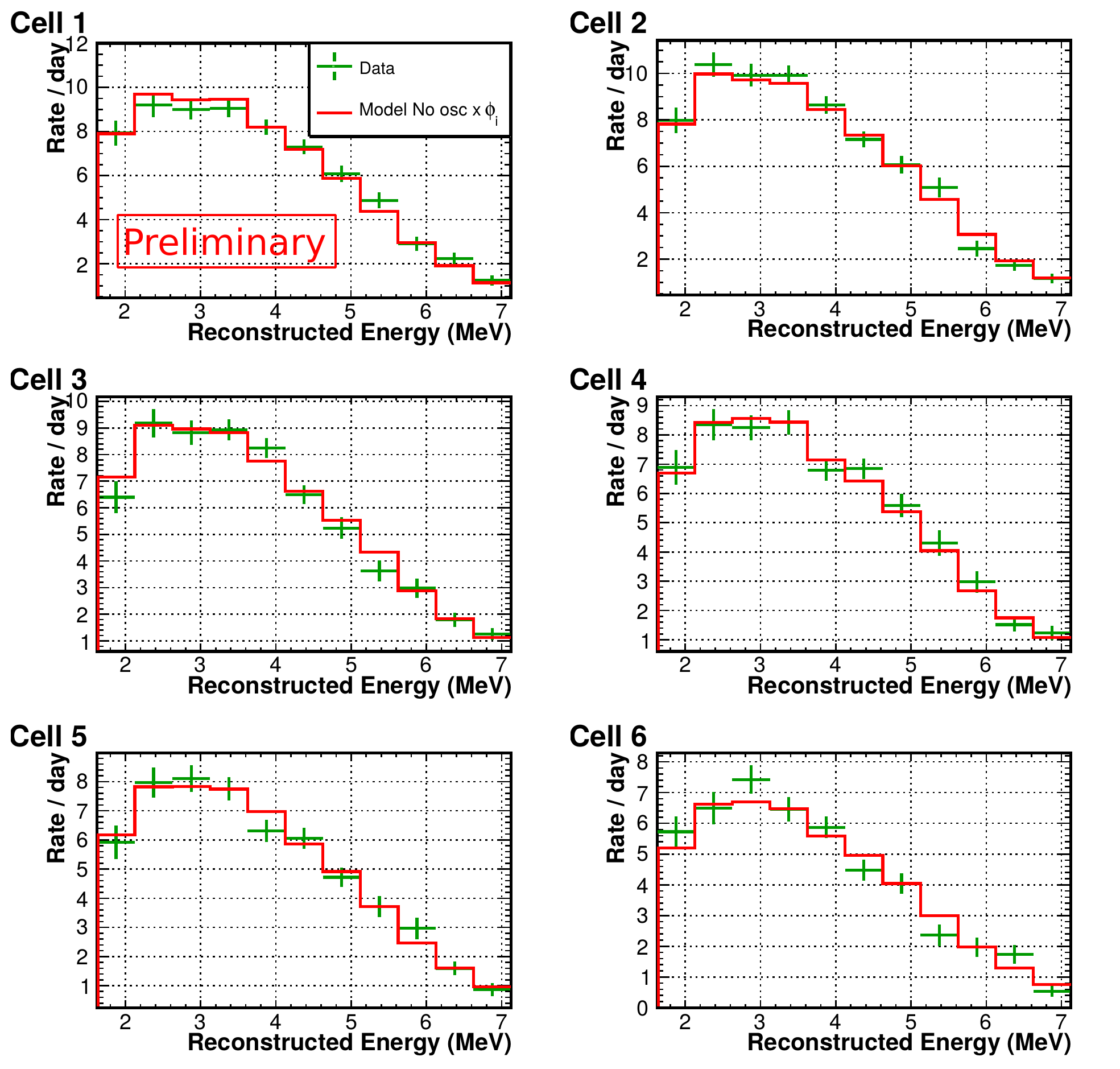}
		\caption{Comparison of the measured neutrino spectra with the non-oscillated model multiplied by the \Phii parameters.}
		\label{fig:Ledieb_Spectra2}
	\end{minipage}%
	\hspace{.2cm}
	\begin{minipage}{0.4\textwidth}
		\centering
		\hspace*{-1.cm}\includegraphics[width=1.16\textwidth]{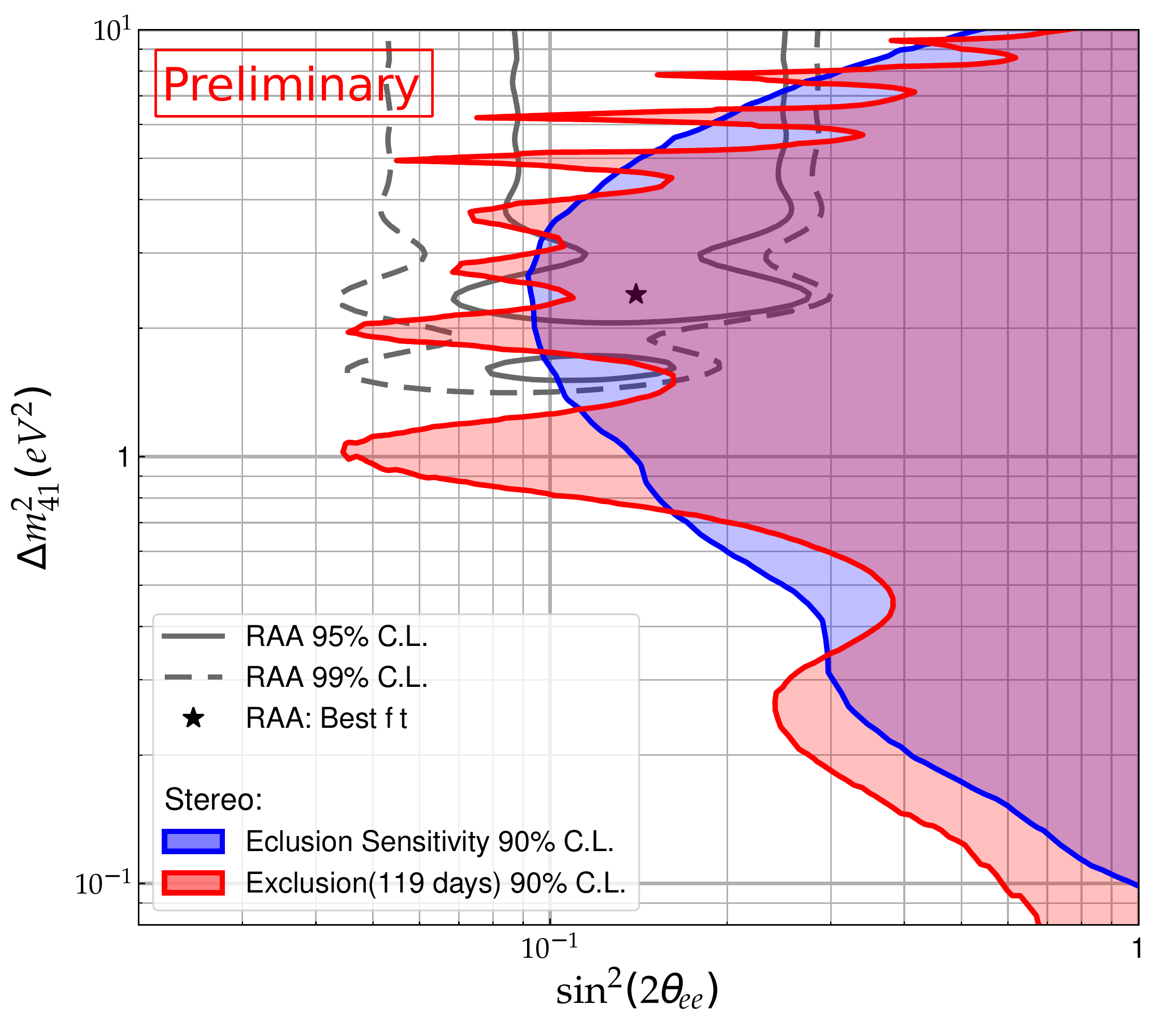}
		\vspace{0.2cm}
		\caption{Exclusion (red) of the sterile neutrino oscillation hypotheses described by \sinDeltam parameters. The sensitivity (blue) is superimposed.}
		\label{fig:ExclusionContours-PhaseII}
	\end{minipage}
\end{figure}

\paragraph{Results}
 The test of the non-oscillation hypothesis gives a p-value of 0.4. It is therefore not rejected. The pull terms show no tension beyond the estimated uncertainties and the measured spectra, displayed in the Figure \ref{fig:Ledieb_Spectra2}, reveal a very good agreement with the non-oscillated models. To produce the exclusion contour (cf. Figure \ref{fig:ExclusionContours-PhaseII}), a raster-scan method is employed and the minimization of each hypothesis is carried out in a fixed $\Delta m^2_{41}$ slice. The statistical fluctuations are well distributed around the expected sensitivity contour. The RAA best-fit value is excluded at $\sim$99\% C.L.

\section{Conclusion}
\paragraph{}
More than 65500 neutrinos have been detected within 119 days of reactor-on using a PSD-based extraction method. Extensive measurements during the reactor-off periods show a high stability of the cosmogenic dominated background. A major fraction of the initial RAA contour is now rejected with no significant cell-to-cell systematics beyond the current statistical fluctuations.

\end{document}